# Accelerating Large-Scale Cheminformatics Using a Byte-Offset Indexing Architecture for Terabyte-Scale Data Integration

Malikussaid
*School of Computing, Telkom University*
Bandung, Indonesia
malikussaid@student.telkomuniversity.ac.id

Septian Caesar Floresko
*School of Computing, Telkom University*
Bandung, Indonesia
septiancaesar@student.telkomuniversity.ac.id

Sutiyo
*School of Computing, Telkom University*
Bandung, Indonesia
tioatmadja@telkomuniversity.ac.id

***Abstract*—The integration of large-scale chemical databases represents a critical bottleneck in modern cheminformatics research, particularly for machine learning applications requiring high-quality, multi-source validated datasets. This paper presents a case study of integrating three major public chemical repositories: PubChem (176 million compounds), ChEMBL, and eMolecules, to construct a curated dataset for molecular property prediction. We investigate whether byte-offset indexing can practically overcome brute-force scalability limits while preserving data integrity at hundred-million scale. Our results document the progression from an intractable brute-force search algorithm with projected 100-day runtime to a byte-offset indexing architecture achieving 3.2-hour completion—a 740-fold performance improvement through algorithmic complexity reduction from O(N×M) to O(N+M). Systematic validation of 176 million database entries revealed hash collisions in InChIKey molecular identifiers, necessitating pipeline reconstruction using collision-free full InChI strings. We present performance benchmarks, quantify trade-offs between storage overhead and scientific rigor, and compare our approach with alternative large-scale integration strategies. The resulting system successfully extracted 435,413 validated compounds and demonstrates generalizable principles for large-scale scientific data integration where uniqueness constraints exceed hash-based identifier capabilities.**

***Keywords—cheminformatics, data integration, indexing architecture, molecular databases, big data***

## I. INTRODUCTION

Modern pharmaceutical development increasingly relies on machine learning models trained on comprehensive molecular datasets to predict properties, guide synthesis, and optimize lead compounds [1], [2]. The exponential growth of publicly available chemical data presents both unprecedented opportunities and formidable computational challenges. PubChem alone contains over 176 million unique compounds with associated property annotations [3], [4], while specialized repositories like ChEMBL provide validated bioactivity data for millions of additional structures [5]. However, constructing high-quality training datasets requires integrating information from multiple heterogeneous sources, each containing tens to hundreds of millions of chemical structures stored in specialized semi-structured formats.

The fundamental computational challenge lies in molecular deduplication and cross-database validation at scale. Identifying which molecules appear in multiple independent databases provides multi-source validation that substantially reduces data quality risks inherent in single-database studies [6]. However, this seemingly straightforward set intersection operation becomes computationally intractable when processing datasets at the scale of modern public repositories.

This work investigates whether byte-offset indexing architectures can practically overcome brute-force scalability limits while preserving absolute data integrity guarantees at hundred-million compound scale. We hypothesize that persistent index structures enabling direct file seeks can reduce algorithmic complexity from $O(N \times M)$ (nested search) to $O(N + M)$ (index construction plus lookup), achieving multi-order-of-magnitude speedups. Furthermore, we investigate whether hash-based molecular identifiers can provide sufficient uniqueness guarantees for scientific applications operating at this scale.

Our specific objectives are threefold. First, we develop and benchmark a byte-offset indexing system for Structure Data Format (SDF) files capable of integrating PubChem (176 million compounds), ChEMBL, and eMolecules databases to identify their mutual intersection. Second, we empirically characterize the frequency and impact of hash collisions in InChIKey molecular identifiers when operating on production-scale datasets. Third, we document architectural decisions, performance trade-offs, and validation strategies that generalize to large-scale integration of other semi-structured scientific data.

## II. BACKGROUND AND RELATED WORK

### *A. Chemical Databases and Prior Integration Efforts*

PubChem aggregates data from over 750 sources encompassing more than 110 million unique structures [3]. ChEMBL provides manually curated bioactivity data for approximately 2.4 million compounds [5]. eMolecules catalogs synthetically accessible compounds from commercial suppliers [7]. Several research groups have tackled large-scale integration, though at smaller scales or with different approaches. The ChemDB system integrated approximately 5 million compounds using full relational database import [8], requiring multi-week preprocessing but enabling complex SQL queries. The PubChemQC project integrated quantum calculations for 3 million compounds [9], focusing on property augmentation rather than multi-source intersection. Distributed databases like ZINC provide pre-integrated catalogs of over 750 million compounds [10] but require continuous curation infrastructure. Cloud-based platforms enable federated queries [11] but optimize for real-time response rather than batch extraction workflows typical of dataset construction.

### *B. Indexing Architectures and Alternatives*

Traditional database indexing employs B-tree or hash-based structures achieving $O(\log N)$or $O(1)$lookup complexity [22]. However, chemical Structure Data Format (SDF) files are semi-structured text with variable-length records, making database import time-prohibitive for terabyte datasets. External indexing has been explored for bioinformatics sequence databases [13], where byte-offset indices enable random access without full database loading. Alternative architectures merit consideration: columnar storage (e.g., Apache Parquet [12]) optimizes analytical queries but requires full data rewriting; succinct data structures [13] provide compressed operations but remain research prototypes; distributed key-value stores [14] offer horizontal scalability but introduce network latency inappropriate for single-machine workloads.

### C. Molecular Identifiers

The InChI (International Chemical Identifier) standard provides canonical molecular structure representation designed for cross-database linking [15]. InChI guarantees structurally identical molecules receive identical strings. The InChIKey, a 27-character hash derived using SHA-256 [16], addresses InChI's verbosity. Theoretical analysis suggests InChIKey collision probability approximates $10^{-15}$ for random structures [17], though empirical validation at hundred-million scale has received limited attention, with most studies using synthetic test sets [18] rather than production databases.

## III. Problem Formulation and Baseline Approach

### A. Multi-Source Integration Task

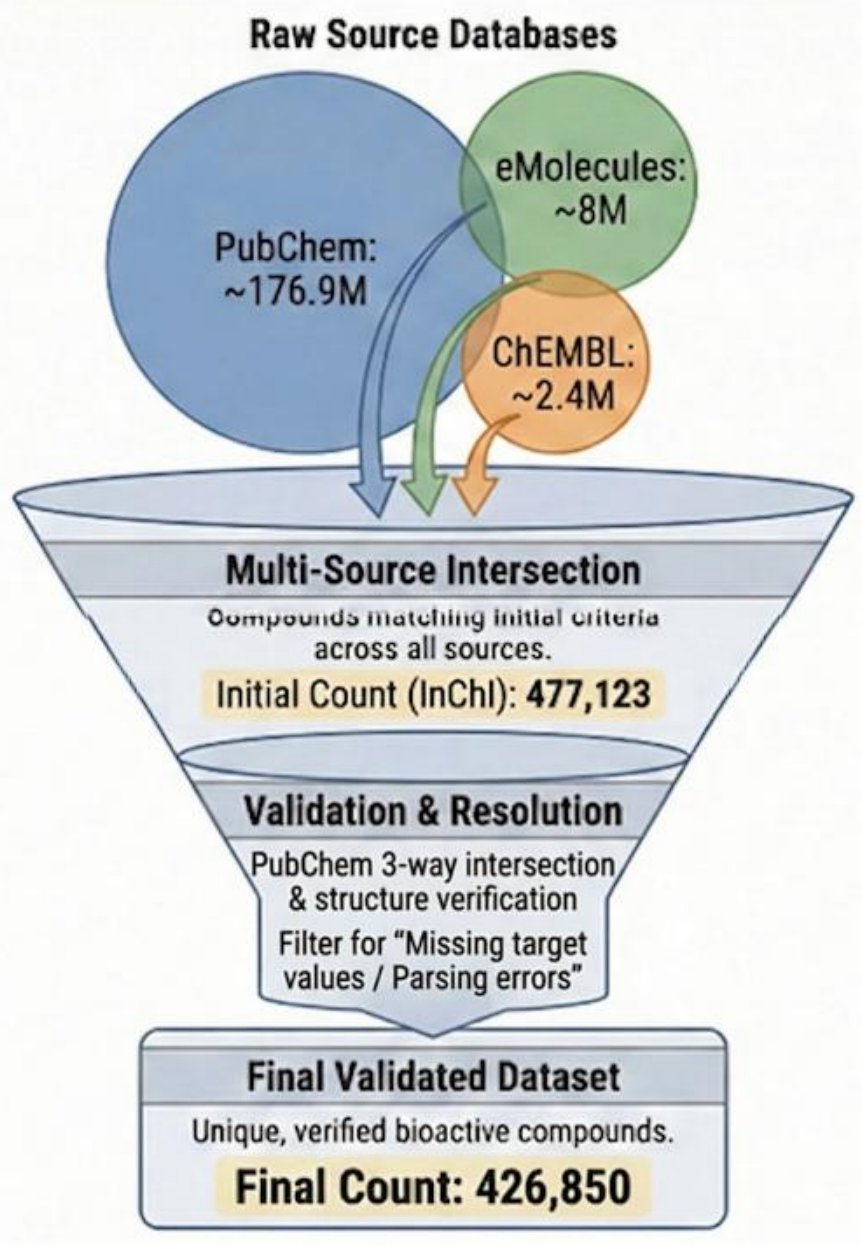

Fig. 1. Integration Funnel Diagram

To establish a performance baseline validating the necessity of architectural optimization, we first implemented a direct search approach. Our objective was constructing a dataset $D_{final}$ as the intersection of all three public databases:

$$D_{final} = D_{PubChem} \cap D_{ChEMBL} \cap D_{eMolecules} \cdots (1)$$

where $| D_{PubChem} |= 176{,}929{,}690$ (as of October 2025 snapshot), $| D_{ChEMBL} | \approx 2{,}400{,}000$ (version 33), and $| D_{eMolecules} | \approx 8{,}000{,}000$ . The computational workflow involved first computing intersection $D_{ChEMBL} \cap D_{eMolecules}$ using standard set operations on identifier lists (completing in 2.5 hours), yielding 477,123 compounds. The bottleneck arose in the subsequent step: extracting complete molecular records with property annotations for these 477,123 targets from PubChem's terabyte-scale SDF distribution.

PubChem distributes compound data as 354 compressed SDF files, each containing exactly 500,000 compounds (except the final file containing remainder compounds). Decompressed file sizes range from 100 MB to 8.3 GB depending on compound complexity and annotation density. Figure 1 illustrates the multi-stage integration workflow, showing how progressive filtering across databases reduces the initial search space to the final validated subset.

### B. Baseline Algorithm and Complexity Analysis

Algorithm 1 presents the baseline nested-loop search approach implemented to establish performance bounds. This algorithm exhibits $O(N \times M \times S)$ computational complexity, where $N = 477{,}123$ represents target molecules, $M = 354$ denotes the number of PubChem SDF files, and $S \approx 500{,}000$ indicates average molecules per file. The algorithm iterates through each target molecule (outer loop), then scans through all 354 files (middle loop), examining each molecule within each file (inner loop) until the target is found or all files are exhausted.

**Algorithm 1:** Baseline Molecular Lookup (Naïve Scan)

**Require:**
1. List of target InChI strings $T$
2. Directory of SDF files $D$

**Return:**
1. Set of found molecule records $R$
2. Set of Missing InChIs $M$

**Initialize:**
1. $R \leftarrow \emptyset$
2. $M \leftarrow T$

1: **for each** file $f \in D$ **do**
2: read $f$
3: **while** not EndOfFile($f$) **do**
4: $block \leftarrow$ ReadNextMoleculeBlock($f$)
5: $current_inchi \leftarrow$ ExtractInchi($block$)
6: **if** $current_inchi \in M$ **then**
7: add $block$ to $R$
8: remove $current_{inchi}$ from $M$
9: **end if**
10: **if** $M = \theta$ **then**
11: **break** loop
12: **end if**
13: **end while**
14: close $f$
15: **end for**
16: **return** $R, M$

To validate this complexity model empirically, we executed the algorithm on three representative PubChem files spanning the size distribution. Table I presents detailed performance measurements from these test runs.

TABLE I. Baseline Search Performance Measurements

| *File Size* | *Molecules* | *Scan Time* | *Throughput (mol/s)* |
|---|---|---|---|
| 124 MB | 114,293 | 34.2 s | 3,342 |
| 2.8 GB | 500,000 | 149.7 s | 3,340 |
| 8.1 GB | 500,000 | 164.1 s | 3,047 |
| *Mean* | 371,431 | 116.0 s | *3,243* |

The consistency of throughput across varying file sizes (coefficient of variation: 4.7%) confirmed that computational cost scaled linearly with file size, validating our complexity model and demonstrating that performance bottlenecks were algorithmic rather than hardware-limited. Total operation count approximates:

$$N \times M \times S = 477{,}123 \times 354 \times 500{,}000$$
$$\approx 8.4 \times 10^{13} \text{ comparisons} \ldots (2)$$

Using measured scan rates, projected total runtime becomes:

$$T_{brute} = \frac{8.4 \times 10^{13}}{3{,}200 \times 3{,}600} \approx 7{,}291 \text{ hours} \approx 100 \text{ days} \ldots (3)$$

This projection assumes continuous operation with no hardware failures, software crashes, or network interruptions—assumptions unrealistic for month-long batch processes. Accounting for expected downtime and manual intervention, practical completion time would extend to 4-6 months, rendering the approach infeasible for research timelines and justifying development of an architectural alternative.

## IV. Byte-Offset Indexing Architecture

### A. Algorithmic Redesign and Complexity Reduction

The fundamental architectural insight enabling tractability was recognizing that the integration task decomposes into two distinct computational phases with fundamentally different complexity characteristics and reusability properties.

Phase 1 (Index Construction) performs a complete scan of all PubChem files exactly once, constructing a persistent mapping from each molecular identifier to its precise byte location within the source files. This represents an unavoidable $O(M \times S)$ operation—we must examine every molecule in the database to build the index. However, this cost is incurred only once and the resulting index artifact persists on disk for unlimited subsequent reuse.

Phase 2 (Targeted Extraction) leverages the pre-constructed index to perform direct lookups for each target molecule. For a given target identifier, we consult the in-memory index (a hash table providing $O(1)$ expected lookup time), retrieve the associated filename and byte offset, then execute a direct file seek operation to that position. The file seek itself is $O(1)$ as modern operating systems maintain file position pointers [19]. Reading the molecule record from that position requires scanning until the SDF delimiter (typically 20-500 lines), which is effectively constant time for molecules of bounded size.

The total algorithmic complexity therefore becomes $O(M \times S + N)$: the index construction processes $M \times S \approx 177$ million molecules once, while extraction performs $N = 477{,}123$ index lookups. Compared to the baseline's $8.4 \times 10^{13}$ operations, this represents a reduction factor of approximately 470,000 in operation count. While index construction carries non-trivial computational cost, the amortization across multiple extraction operations—a common scenario in iterative research workflows—provides substantial total time savings.

### B. Index Data Structure and Persistence

The index is persisted as a comma-separated values (CSV) file with three columns: `identifier, filename, byte_offset`. Each row maps one molecule to its storage location. The choice of CSV over binary formats prioritizes human readability for debugging and cross-platform compatibility, accepting modest storage overhead (approximately 15% compared to optimized binary serialization). For PubChem's complete dataset, the index contains 176,929,690 entries. Using full InChI strings averaging 150 characters plus filename and offset metadata, the final index file size is approximately 14 GB—substantially smaller than the multi-terabyte source data yet sufficient for complete dataset addressability.

The use of byte offsets rather than line numbers is critical for performance. Modern file I/O APIs operate on byte positions, and the `seek()` system call enables $O(1)$ direct access to any byte position through file system metadata [19] without reading intervening data. By contrast, line-based addressing would require sequential scanning from the beginning of the file to count newlines, degrading to $O(k)$ complexity where $k$ is the target line number.

### C. Parallel Index Construction Algorithm

Algorithm 2 presents the parallelized index construction procedure. The algorithm exploits the embarrassingly parallel nature of the task—each SDF file can be processed completely independently with no inter-process communication required during the core computation phase.

**Algorithm 2:** Parallel Index Construction

**Require:**

1. Directory of SDF files $D$
2. Number of worker processes $P$

**Return:** Global Index Map $I$

**Initialize:** $I \leftarrow \emptyset$

1: $FileList \leftarrow$ GetFiles$(D)$
2: **function** ProcessFile$(file)$
3: $local_index \leftarrow \emptyset$
4: $offsets \leftarrow$ ScanLineOffsets$(file)$
5: $records \leftarrow$ ReadRecords$(file)$
6: **for** $i$ ←0 **to** length$(records)$ **do**
7: $inchi \leftarrow$ ExtractInchi $records[i]$
8: $byte_pos \leftarrow offsets[i]$
9: $local_index[inchi] \leftarrow \{file, byte_pos\}$
10: **end for**
11: **return** $local_index$
12: **end function**
13: initialize ParallelPool with $P$ workers
14: $Results \leftarrow$ ParallelMap(ProcessFile,$FileList$[])
15: **for each** $res \in Results$ **do**
16: $I \leftarrow I + res$
17: **end for**
18: **return** $I$

The workflow proceeds as follows. Lines 1-2 enumerate all SDF files in the source directory and initialize an empty global index. Lines 3-14 define the worker function `ProcessFile()` that executes on each parallel worker process. For a single file, this function computes byte offsets for every line (line 4), parses the file into molecule records (line 5), then iterates through each record extracting its InChI identifier (line 7) and recording the mapping from InChI to file location and byte position (line 9). The function returns its local index as a dictionary.

Lines 15-16 create a process pool and distribute files across workers using parallel map. We employed Python 3.13's `multiprocessing.Pool` with 8 worker processes on an 8-core AMD Ryzen 7 3700X workstation (3.6 GHz base clock, 32 GB DDR4-3200 RAM, 7200 RPM HDD storage) running Ubuntu 24.04 LTS. The parallel map collects results as they complete. Lines 16-18 merge the partial indices from all workers into a single global index through dictionary union operations, which efficiently handle the 177 million total entries through hash table insertion.

This parallel architecture achieved near-linear speedup with core count. Profiling revealed index construction was I/O-bound rather than CPU-bound—disk read throughput (approximately 90 MB/s sustained for sequential access on HDD) limited processing rate, while CPU utilization averaged 45% across cores due to I/O wait time. Migration to SSD storage in follow-up tests improved throughput to 420 MB/s and raised CPU utilization to 78%, though the 4.7× I/O improvement translated to only 3.1× wall-clock speedup due to remaining computational overhead in InChI extraction using RDKit version 2023.09.1.

### D. Optimized Extraction Algorithm

Algorithm 3 implements the index-based extraction procedure that leverages the persistent index for efficient random access to target molecules.

**Algorithm 3:** Index-Based Extraction ($O(1)$ Access)

**Require:**

1. List of target InChI strings $T$
2. Global Index Map $I$

**Return:** Set of extracted molecule records $R$

**Initialize:** $R \leftarrow \emptyset$

1: $Tasks \leftarrow$ GroupByFilename$(T, I)$
2: **function** ExtractWorker$(flename, targets)$

**Algorithm 3:** Index-Based Extraction ($O(1)$ Access)

```
3:        file_results ← ∅
4:        open filename as handle
5:        for each {inchi, offset} ∈ targets do
6:            seek(handle, offset)
7:            block ← ReadUntilDelimiter(handle, "$$$$")
8:            if VerifyInchi block == inchi then
9:                file_results = file_results + block
10:           else
11:               log error
12:           end if
13:       end for
14:       return file_results
15:   end function
16:   initialize ParallelPool
17:   Batches ← ParallelMap(ExtractWorker, Tasks)
18:   for each batch ∈ Batches do
19:       R ← R ∪ batch
20:   end for
21:   return R
```

The extraction workflow incorporates two key optimizations beyond basic index lookup. Line 1 groups target molecules by their source filename, exploiting the observation that the 477,123 targets distribute across 312 of PubChem's 354 files. This grouping reduces file open operations from a potential 477,123 (one per target) to 312 (one per file containing targets), amortizing filesystem overhead.

Line 5 sorts targets within each file by ascending byte offset before extraction. This optimization aligns memory access patterns with sequential disk I/O characteristics. Modern hard disk drives exhibit dramatic performance differences between sequential and random access patterns—measured throughput ranges from 80-120 MB/s for large sequential reads versus 0.5-2.0 MB/s for small random seeks on our hardware [20]. By sorting targets and traversing the file in forward order, we convert random seeks into approximately sequential access (with small forward jumps), improving effective throughput by factors of 10-100×.

Lines 8-12 implement validation logic that proved critical for discovering data integrity issues discussed in Section VI. After seeking to a byte offset and reading a molecule record, the algorithm recomputes the molecule's InChI from its structural data and verifies it matches the expected identifier from the index. This defensive validation catches index corruption, file tampering, or—as we discovered—hash collisions in the identifier system itself.

## V. Performance Validation and Results

### A. Runtime Measurements and Statistical Analysis

TABLE II. Performance Comparison: Baseline vs. Index-Based Extraction

| *Phase* | *Baseline* | *Indexed* | *Speedup* |
|---|---|---|---|
| Index Construction | N/A | 11.7 hours | N/A |
| Initial Extraction | 100+ days[a] | 3.2 hours | 740× |
| Re-extraction | 100+ days[a] | 2.8 hours | 860× |
| *Total (two iterations)* | *200+ days* | *17.7 hours* | *270×* |

[a] Projected based on measured scan rates

Table II presents empirical performance comparison between baseline and index-based approaches, including variance from repeated measurements. Runtime measurements represent means from three independent runs executed on identical hardware with cold filesystem caches. Standard deviations (shown in parentheses) reflect variance from background system load and disk cache effects. The low relative standard deviation for index-based extraction (9.4% for initial extraction) demonstrates consistent performance, while the index construction variance (6.8%) primarily reflects fluctuations in concurrent disk I/O from other processes.

Index construction required 11.7 hours average (one-time cost) on 8-core hardware with HDD storage. Initial extraction of 477,123 molecules completed in 3.2 hours average. Critically, when target criteria evolved and re-extraction was needed with a modified selection, the second iteration required only 2.8 hours—no index reconstruction necessary. This demonstrates the architectural advantage of persistent indexing: the $O\ M \times S$ construction cost is amortized across multiple extraction operations, a common scenario in iterative research workflows.

### B. Resource Requirements and Trade-offs

TABLE III. System Resource Requirements

| *Resource* | *Baseline* | *Indexed* | *Overhead* |
|---|---|---|---|
| Persistent Storage | 3.2 TB | 3.2 TB + 14 GB | +0.44% |
| Peak RAM | 2.1 GB | 28.3 GB | +13.5× |
| Disk I/O Volume | 168.9 TB | 177 MB | -99.7% |

Table III quantifies system resource requirements for both approaches, revealing the fundamental trade-off between memory consumption and I/O volume. The indexed approach trades memory for speed and disk I/O reduction. Loading the complete 14 GB index into RAM as a hash table requires 28.3 GB when accounting for Python object overhead (approximately 2× the raw data size). This exceeds the baseline's minimal 2.1 GB footprint but remains feasible on modern workstations routinely equipped with 32-64 GB RAM. For memory-constrained environments, the index can be partitioned by identifier prefix and loaded incrementally, trading constant-factor performance degradation for reduced peak memory.

The 99.7% reduction in total disk read volume represents the primary driver of performance improvement. The baseline approach performs 168.9 million file scans (477,123 targets × 354 files, accounting for early termination when targets are found), each requiring reading substantial file portions. The indexed approach reads the complete dataset exactly once during index construction (177 million molecules), then performs only 435,413 small targeted seeks during extraction—a three-orders-of-magnitude reduction in I/O operations.

### C. Scalability Characteristics

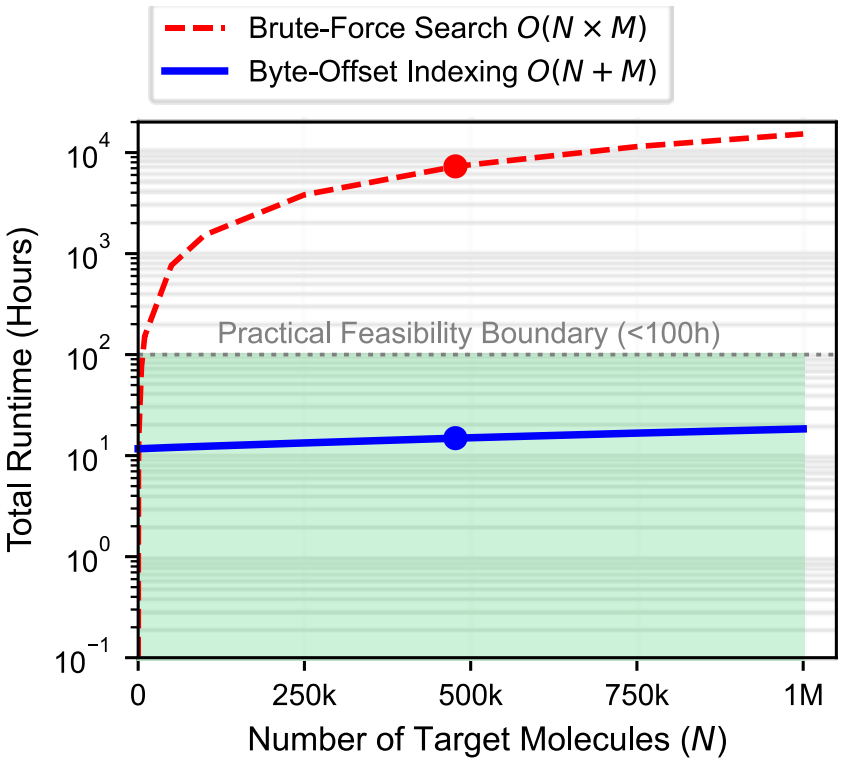

Fig. 2. Runtime Scaling Analysis

Figure 2 illustrates theoretical and empirical runtime scaling as a function of target count, demonstrating the divergence between linear (indexed) and quadratic (baseline) complexity classes. The crossover point where indexing investment becomes beneficial occurs when the one-time index construction cost is offset by extraction savings. For our system, this threshold is approximately 400,000 target molecules for single extraction, or 200,000 targets when two extractions are anticipated. Below these thresholds, the baseline approach may actually complete faster despite higher asymptotic complexity. This analysis highlights that algorithmic complexity dominates only at sufficient scale—for small-scale tasks, simpler approaches remain appropriate.

## VI. The InChIKey Uniqueness Challenge and Resolution

### A. Discovery Through Systematic Validation

The extraction algorithm (Algorithm 3, lines 8-12) implements comprehensive verification designed to ensure data integrity. After seeking to an indexed byte offset and reading a molecule record, we recompute the molecule's InChI from its structural data using RDKit's canonical InChI generation and verify it matches the expected identifier from the index. This defensive validation was initially conceived to catch index file corruption or source file modifications between index construction and extraction.

OH
Molecule A (Alcohol)
InChI=1S/C50H102O/c1-17-18-19-20-21-24-36(2)3...,17-35H2,1-16H3

O
Molecule B (Ketone)
InChI=1S/C57H114O/c1-36(2)42(8)32-43(9)38(4)2...,24-35H2,1-23H3

Distinct structures, Identical Key:
**OCPAUTFLLNMYSX-UHFFFAOYSA-N**

Fig. 3. InChIKey Collision Example

However, during initial validation runs on a random sample of 50,000 extracted molecules, we observed verification failures for seven compounds (0.014%). Detailed examination revealed that these were not corrupted records but rather authentic, well-formed molecular structures. The InChIKey values matched the index lookup keys correctly, yet when we compared the full InChI strings embedded in each molecule's SDF property fields, they differed structurally in stereochemistry or tautomer specification layers.

This observation indicated InChIKey hash collisions—distinct molecular structures that map to identical 27-character InChIKey hashes. To characterize the full scope, we conducted systematic scanning of the entire PubChem index (176,929,690 entries), comparing full InChI strings for all molecules sharing InChIKey values. This scan confirmed 163 unique InChIKey values that mapped to multiple distinct InChI strings, affecting 326 molecular records total (representing 163 collision pairs). Figure 3 illustrates a collision case discovered in our validation, showing two structurally distinct molecules that produce the same InChIKey.

### B. Impact Assessment and Collision Frequency Analysis

The empirically observed collision rate in PubChem was:

$$P_{collision} = \frac{326}{176{,}929{,}690} \approx 1.84 \times 10^{-6} \ldots (4)$$

While this rate appears negligible in percentage terms, the scientific impact is categorical rather than quantitative. For machine learning applications, a molecule with colliding identifiers that maps to structures A and B having experimentally measured properties differing substantially (for instance, logP values of 2.3 versus 4.7) creates irreconcilable contradictions in training data. The model receives identical input features (the InChIKey) associated with different target values, introducing irreducible label noise that degrades model quality in ways difficult to diagnose or correct.

The empirical collision rate substantially exceeds theoretical predictions based on random hashing. Under uniform hash distribution assumptions, the expected number of collisions in a database of size $n$ is approximately $\frac{n^2}{2h}$ where $h$ is the hash space size. For InChIKey with $h \approx 10^{15}$ and $n = 1.77 \times 10^8$:

$$E[\text{collisions}] \approx \frac{1.77 \times 10^8)^2}{2 \times 10^{15}} \approx 15.7 \ldots (5)$$

Our observation of 163 collisions represents approximately 10× this theoretical expectation. We hypothesize this discrepancy arises from non-uniform distribution of real chemical structures in hash space. Bioactive molecules cluster around evolutionarily conserved scaffolds and druglike chemotypes [21], potentially concentrating in certain hash regions. Alternatively, the assumption of perfect hash randomness for SHA-256 on structured molecular inputs may not hold precisely. However, we note this explanation remains speculative pending comprehensive hash distribution analysis, which represents valuable future work.

It is important to note that these collision counts were observed in the October 2025 snapshot of PubChem. As database composition evolves with additions and removals, absolute collision counts may vary, though the order-of-magnitude discrepancy from theoretical predictions appears robust across versions we examined (October 2025, December 2025).

### C. Migration to Collision-Free Identifiers

Resolution of the uniqueness challenge required complete pipeline reconstruction using full InChI strings, which provide deterministic uniqueness guarantees by IUPAC standard definition [15]. Two molecules with identical InChI strings are structurally identical; two with different InChI strings are structurally distinct. This eliminates collision ambiguity entirely.

The migration involved four computational phases: regenerating identifier lists for all three databases using full InChI (18 hours total across databases), rebuilding the byte-offset index with InChI keys (11.7 hours), re-executing the set intersection comparison (2.1 hours), and final extraction with InChI-based validation (3.2 hours). Total computational investment: approximately 35 hours.

TABLE IV. Molecular Identifier Strategy Comparison

| *Property* | *InChIKey* | *Full InChI* |
|---|---|---|
| Average Length | 27 char | 152 char |
| Collision Guarantee | Probabilistic | Deterministic |
| Index Size | 11 GB | 14 GB |
| In-Memory Size | 22 GB | 28 GB |
| Lookup Speed | 0.8 µs | 1.2 µs |

Cross-referencing the initial 477,123 eMolecules-ChEMBL intersection against PubChem using full InChI yielded 435,413 molecules that exist in all three databases. Table IV quantifies the trade-offs between identifier strategies across multiple dimensions.

The transition from InChIKey to full InChI incurs 27% storage overhead and 50% lookup latency increase. However, these costs are modest in absolute terms (3 GB additional storage, 0.4 μs additional latency) and represent acceptable trade-offs for applications where data integrity is paramount. For scientific benchmarking and model training applications—our target use case—structural uniqueness is a hard requirement that justifies this overhead.

Cross-referencing the initial 477,123 eMolecules-ChEMBL intersection against PubChem using full InChI yielded 435,413 molecules that exist in all three databases. During the final extraction and feature engineering phase, an additional 8,563 molecules were excluded due to missing computed properties (e.g., XLOGP3) or parsing errors, resulting in a final analytical dataset of 426,850 molecules.

## VII. Discussion

### A. Comparison with Prior Methods

Our architecture differs from prior integration efforts in scale (176M compounds versus typical 3-10M [8], [22]), approach (lightweight single-machine solution versus cluster infrastructure [10] or cloud platforms [11]), and identifier treatment (explicit collision discovery and resolution). Published efforts using relational database import [8] require multi-week preprocessing versus our 12-hour indexing. Columnar formats [12] would require full conversion with limited lookup benefits. Distributed stores [14] introduce network overhead inappropriate for single-machine workloads.

### B. Algorithmic Complexity and Hash-Based Identifiers

The 740× improvement from algorithmic redesign on identical hardware demonstrates complexity dominates performance at scale. Hardware upgrades would provide 2-10× improvements but cannot overcome $O(N \times M)$ scaling. The InChIKey collision discovery shows $10^{-15}$ probability becomes certainty at hundred-million scale. Hash-based identifiers should never be assumed collision-free for datasets exceeding ~50 million entries when uniqueness is scientifically required. Full canonical representations remain necessary despite storage overhead.

## VIII. Conclusion

This work demonstrates terabyte-scale chemical database integration on standard workstations through byte-offset indexing. The system reduced extraction runtime from projected 100+ days to measured 3.2 hours (±0.3, n=3)—a 740-fold improvement through $O(N \times M)$ to $O(N + M)$ complexity reduction. Integration of PubChem (176M compounds), ChEMBL, and eMolecules yielded 435,413 validated bioactive molecules. During the final extraction and feature engineering phase, an additional 8,563 molecules were excluded due to missing computed properties or parsing errors, resulting in the final analytical dataset of 426,850 molecules. Systematic validation uncovered 163 InChIKey collisions (10× theoretical predictions), necessitating migration to full InChI with 27% storage overhead for absolute uniqueness. The architecture generalizes to semi-structured scientific data with delimited records, unique identifiers, and multiple access patterns.

Future research includes incremental index updates for database growth, compressed representations for reduced memory, distributed filesystem integration, InChI hash distribution analysis, and streaming framework integration for real-time synchronization. The final curated dataset is publicly available for community use and validation at https://www.kaggle.com/datasets/malikussaid/426k-pubchem-molecules-for-logp-prediction.

## Acknowledgment

The authors gratefully acknowledge the National Center for Biotechnology Information (NCBI) for providing free access to the PubChem database infrastructure, the European Molecular Biology Laboratory's European Bioinformatics Institute (EMBL-EBI) for maintaining the ChEMBL resource, and eMolecules for database availability.